# Dual-Model Prediction of Affective Engagement and Vocal Attractiveness from Speaker Expressiveness in Video Learning

Hung-Yue Suen, Kuo-En Hung, and Fan-Hsun Tseng, *Senior Member, IEEE*

*Abstract*—This paper outlines a machine learning-enabled speaker-centric Emotion AI approach capable of predicting audience-affective engagement and vocal attractiveness in asynchronous video-based learning, relying solely on speaker-side affective expressions. Inspired by the demand for scalable, privacy-preserving affective computing applications, this speaker-centric Emotion AI approach incorporates two distinct regression models that leverage a massive corpus developed within Massive Open Online Courses (MOOCs) to enable affectively engaging experiences. The regression model predicting affective engagement is developed by assimilating emotional expressions emanating from facial dynamics, oculomotor features, prosody, and cognitive semantics, while incorporating a second regression model to predict vocal attractiveness based exclusively on speaker-side acoustic features. Notably, on speaker-independent test sets, both regression models yielded impressive predictive performance ($R^2 = 0.85$ for affective engagement and $R^2 = 0.88$ for vocal attractiveness), confirming that speaker-side affect can functionally represent aggregated audience feedback. This paper provides a speaker-centric Emotion AI approach substantiated by an empirical study discovering that speaker-side multimodal features, including acoustics, can prospectively forecast audience feedback without necessarily employing audience-side input information.

## I. INTRODUCTION

Asynchronous video-based instruction, accelerated by the expansion of MOOC platforms and digital learning technologies, has redefined how educational and professional content is disseminated at scale [1]. While offering convenience and broad accessibility, this format introduces a pedagogical limitation: the difficulty of maintaining affective engagement, particularly in the absence of real-time learner feedback. Affective engagement can be characterized by emotional resonance, sustained attention, and motivational alignment. Such engagement has been shown to influence learning persistence and academic outcomes [2], [3].

Recent research indicates that speaker expressiveness encompasses facial dynamics, vocal prosody, and emotionally charged words. Speaker expression is associated with sustaining audience engagement, especially within asynchronous, non-interactive environments [4], [5]. However, most existing engagement assessment methods rely on self-report surveys, interaction logs, or biometric sensing, all of which are hindered by limited scalability, delayed feedback, and privacy-related constraints, particularly within asynchronous, large-scale learning settings [6]. In response, recent advances in affective computing have employed deep learning (DL) techniques to estimate engagement from facial expression recognition (FER) and vision-based affect modeling [7], [8]. Yet these approaches assume persistent facial visibility, a condition seldom met in real-world scenarios such as MOOCs or social media-based learning, where viewers participate privately and often without enabling camera access due to technical limitations or regulatory restrictions [9], [10].

To address the limitations of audience-dependent engagement inference, this paper proposes a speaker-side, content-based framework that predicts affective engagement using expressive cues captured solely from the speaker. These include facial movements (e.g., expressions, head orientation, eye motion), vocal prosody, and linguistic content extracted directly from instructional videos. Prior evidence indicates the predictive power of these modalities particularly vocal attractiveness, defined by prosodic variation, clarity, and expressive tone can be regarded as a salient indicator of emotional resonance and perceived engagement [5]. Gaze dynamics, such as fixation stability and saccadic behavior, further reflect cognitive alignment and speaker intent [11]. Nevertheless, reliably quantifying vocal appeal in automated systems remains technically challenging.

Building on these insights, this paper makes two primary contributions. First, we provide a strong empirical demonstration that exclusively speaker-side expressive cues can robustly predict audience-perceived affective engagement and vocal attractiveness in a scalable, privacy-preserving manner. Our findings reveal that acoustic features are the most powerful single predictor. Second, we present a practical, dual-model machine learning (ML) framework that leverages this finding. The framework is designed for real-world deployment,

This work was supported by the National Science and Technology Council (NSTC), Taiwan, under Grants NSTC 112-2410-H-003-102-MY2, NSTC 113-2622-H-003-003, and 114-2410-H-003-019-MY2. All authors contributed equally to this work. (*Corresponding author: Fan-Hsun Tseng.*)

Hung-Yue Suen is with Technology Application and Human Resource Development, National Taiwan Normal University, Taipei, Taiwan (ORCID: 0000-0002-6796-2031; e-mail: collin.suen@ntnu.edu.tw).

Kuo-En Hung is with Technology Application and Human Resource Development, National Taiwan Normal University, Taipei, Taiwan (ORCID: 0000-0003-2091-2747; e-mail: kuanntw@gmail.com).

Fan-Hsun Tseng is with the Department of Computer Science and Information Engineering, National Cheng Kung University, Tainan 701, Taiwan (ORCID: 0000-0003-2461-8377; e-mail: tsengfh@gs.ncku.edu.tw).

offering a modular and interpretable solution for inferring affective outcomes without requiring any viewer-side data, thereby aligning with the principles of data minimization and privacy-by-design found in regulations such as the EU AI Act [12]. It integrates features from facial dynamics, oculomotor, acoustics, and textual (semantic) signals into an optimized ensemble regressor, showcasing a path toward ethically aligned Emotion AI in education.

This paper contributes to affective computing in computational social systems by introducing a dual-model architecture that captures both expressive richness and perceptual salience. Rather than treating affective engagement as a singular latent construct, the framework decomposes it into interpretable dimensions, such as vocal appeal and multimodal expressiveness, each addressed through dedicated learning pathways for enhanced robustness under diverse input conditions.

The proposed modularity supports the deployment of Emotion AI across diverse domains, including digital learning, media communication, and speaker analytics. By relying exclusively on speaker behavior, the framework enables privacy-preserving and user-aware affective inference aligned with principles of transparency and consent. This paper treats affective engagement and vocal attractiveness as distinct yet interrelated outcomes, establishing the latter as a valid proxy for audience-perceived affect. The dual-model design enhances interpretability and resilience in audio-only or privacy-sensitive settings, thereby broadening its applicability to domains such as teleconsultation, virtual coaching, and large-scale content delivery platforms. The main contributions of the article are listed as follows.

1) A dual-model framework for speaker-side affective inference, including a multimodal engagement model and an acoustic-only vocal attractiveness model.
2) Empirical evidence that acoustic features alone can predict audience engagement, supporting their standalone utility when visual or textual data are missing.
3) A privacy-preserving machine learning design that uses only de-identified speaker cues, compliant with standards such as the EU AI Act.
4) A flexible fusion strategy that combines facial motion, oculomotor features, vocal prosody, and semantic content through feature concatenation.
5) A practical foundation for applications in emotion-aware education, coaching, and computational social systems.

The rest of this article is organized as follows. Section II discusses the related works and background. In Section III, the methodology of the proposed prediction framework is presented. Results and evaluations are shown in Section IV. Section V concludes this work.

## II. RELATED WORK AND BACKGROUND

With ongoing developments in engagement modeling, there has been increasing interest in predicting audience affect during video-based learning mediums. Traditional approaches to engagement measurement, such as self-reporting, interactive behavior, and biometrics (e.g., facial expression, eye and head movement), continue to operate within synchronous and controlled settings exclusively [13]. This is because, within asynchronous video-based mediums, there is a lack of immediate access to information concerning learners' behavior, which is a source of growing interest concerning speaker-side predictions via expression [14]. This serves to promote approaches centered on content, which are necessarily scalable and privacy-respecting, with speaker expression being a chief indicator within digital communication applications [15].

### A. Speaker Expressiveness and Affective Engagement

Affective engagement: This term is described as the involvement of the affective state during the consumption of educational content [16], and this is extremely important when aiming to achieve and maintain these engagement outcomes within asynchronous video. As already stated within related literature, according to emotional contagion theory, affect can indeed be triggered by observation of other people's expressive behavior, such as vocal expressions, facial expressions, and emotionally loaded verbal text [17]. The lack of interaction within this scenario forces these speaker-driven expressions to function solely as the main mechanism by which affect is transmitted and influences how these messages resonate affectively with both information and speaker alike [18]. This can indeed be observed within perspectives such as audience design theory [19] and hedonic motivation theories [20].

Empirical work has shown that speaker expressiveness is a predictor of affective engagement [21]. Expressive signals such as facial expressions, gaze behavior, vocal prosody, and affective words convey emotional intent and are predictive of perceived emotional resonance among viewers [15]. Emphasis on voice, affectively framed communication, and speaker eye-gaze have been observed to increase affective engagement and affective connection with viewers [22], [23]. This work provides a strong theoretical and empirical basis to model affective engagement via speaker-side expressive features.

### B. Multimodal Cues for Affective Engagement

Owing to the verified connection between speaker expressiveness and affective engagement with audience members [4], [5], early attempts on multimodal affect recognition have concentrated on unimodal approaches analyzing visual, acoustical, and linguistic information independently. Visual-based approaches tend to make use of convolutional neural networks (CNN) or 3D CNN to identify spatiotemporal information conveyed by facial expressions and head nodding [24]. Acoustic approaches make use of prosody features like pitch, MFCC, and intensity analyzed via long short-term memory (LSTM) networks or transformer encoding to assess valence/arousal [25]. The linguistic approach generally uses sentiment lexicons or embeddings like Bidirectional Encoder Representations from Transformers (BERT) to assess speaker emotion on transcribed speech texts [26].

Though very useful within limited domains, mono-modal approaches are very sensitive to noise, inadequately informative semantically, and generally inefficient within asynchronous and/or true-world applications [27]. These constraints have triggered interest and work on multimodal fusion approaches to leverage complementary information sources across these aspects [28]. Going beyond facial expression, more recent work underlines the need to incorporate oculomotor processes like fixation duration and saccadic eye movement, which serve as pointers to cognitive engagement and attention alignment, reflecting speaker intention and cognitive information processing on a level more nuanced than mere affect expression [11].

Among multimodal approaches, architectures employing deep learning-based fusion have shown considerable success within affective computing applications [29], particularly those incorporating LSTM and transformer architectures. Yet, these architectures tend to work best on relatively large annotated datasets and may face issues such as overfitting within more sparse domains.

To overcome these issues, ensemble approaches like Extreme Gradient Boosting (XGBoost) provide an alternative method to achieve high prediction precision while simultaneously improving interpretability and robustness to heterogeneous and incompletely observed data [30]. Within this paper, we have utilized XGBoost [31] employing Bayesian Optimization to optimize its hyperparameters.

*C. Modeling Speaker Vocal Attractiveness*

Although these multimodal approaches assess a wide range of expressive behaviors, there is more recent evidence to show that purely vocal communications contain potent affective information. These include speaker vocal attractiveness, which has been observed to play a relevant role as a perceptual predictor of audience engagement within video-based environments for learning [5]. A range of important acoustic features associated with speaker vocal attractiveness has been previously identified to contribute to speaker vocal warmth, clarity, and attractiveness, including those related to F0, HNR, jitter, shimmer, energy contours, MFCC values, and formant values [32]–[35]. The role of voice perception similarity on speaker vocal attractiveness has also been identified to play an important part [36].

Current approaches to modelling vocal attractiveness from acoustic input have been limited to controlled environments such as short speech conversations and political speeches. Models using regression have been deployed within limited communication scenarios [37], while those using neural networks have been employed to directly estimate speaker attractiveness from short audio clips [35]. Notably, though there is increased interest in affect modelling using voice, there is no work on modelling speaker vocal attractiveness within asynchronous learning environments.

To fill this research gap, this study aims to design and test a regression model that predicts vocal attractiveness on the basis of acoustic attributes. This regression model can complement the multimodal predictor of engagement to serve both theoretical and practical functions toward affect inference on the speaker-side within e-learning applications.

III. METHODOLOGY

We present a prediction system with speaker-side expressive features to estimate affective engagement and vocal attractiveness according to audience perception within asynchronous video-based learning environments. The proposed system consists of two parallel regression models using XGBoost. The first regression model estimates affective engagement by combining facial, acoustic, and linguistic features, while the second model estimates vocal attractiveness employing acoustic features only. Both regression models generate predictions on a continuous scale corresponding to the 1-5 points on the Likert scale required in post-video surveys. This two-model design leans on recent breakthroughs within multimodal affective computing [8], speaker-aware sentiment analytics [38], and end-to-end social perception architectures [39].

As shown in **Fig. 1**, each input video segment is passed through three modality-specific streams: audio, visual, and transcript. These streams involve preprocessing and extracting features. Acoustic features are extracted from a prosodically stable 2-second segment, encompassing both prosodic and spectral properties. The visual features are extracted over a fixed 2-second window (60 frames) and flattened into a fixed-size vector to retain frame-specific motion information. Oculomotor features obtained from eye-gaze information are added to retain speaker attention information. The features are obtained from automatic transcripts for the text modality. All features are independently extracted and z-normalized before concatenation for regression modeling.

Affective engagement modeling is performed on a concatenated representation formed by combining standardized vectors across all three modalities, which is passed to an XGBoost regressor. For vocal attractiveness modeling, on the other hand, training is performed separately, employing solely acoustic features. This helps to improve interpretability while maintaining modularity rather than using more complex merging approaches. This approach helps to cope flexibly with differing availability across multiple modalities, such as in an audio-only environment.

*A. Data Collection*

The videos are asynchronous instructional videos, which were gathered using Taiwan's top three MOOC platforms. The final dataset comprises 10,360 segments that contain valid facial, acoustic, and text features to train the Affective Engagement model and 9,960 audio-qualified segments to train the Vocal Attractiveness model after filtering on system compatibility.

To support speaker-level generalization and avoid leakage, speaker-independent partitioning is performed. This entails assigning each utterance fragment belonging to a specific teacher to only one part, namely training (70%), validation (15%), and test sets (15%).

Data collection occurred after formal memoranda of understanding and confidentiality agreements were reached with each platform. To ensure compliance with data protection regulations and each platform's specific privacy policies, anonymity was ensured on each sample dataset imported. Only gender information on instructors and domain information on courses remained available, while no personal information on any learners could be obtained. The videos sent to each platform lasted 7-15 minutes (M = 13.62, standard deviation (SD) = 4.51) and involved on-camera presentations. The study's protocol was approved by the authors' Institutional Review Board.

The speakers comprised 72% males and 28% females. The domains were dominated by social sciences (82%) and science domains (18%). The type of conference call selected ensured a sufficient level of audio signal and facial expression quality. To verify how well each model generalizes, initial subgroup analyses were obtained on speaker sex, domains, and length quartiles, employing both the coefficient of determination ($R^2$) and root mean squared error (RMSE) metrics. The results revealed no statistically significant differences in how well each model performed on these subgroups.

The post-class user ratings were used to provide ground truth for supervised learning. Affective engagement was assessed via a five-point scale developed by adapting the Online Student Engagement Scale [40], while vocal attractiveness was assessed via a two-point scale developed by adapting Yuan et al. [41]. The engagement and vocal attractiveness items both consisted of a 1-to-5 Likert scale. The raters were actual learners participating in the MOOC corresponding to each course and asked to rate immediately after each video segment. Only those segments with more than two ratings were included. The distribution of scores approached a Gaussian distribution, meaning these variables were slightly skewed. The affective engagement scores' mean value is 4.079 (SD = 0.441), and the vocal attraction scores' 4.042 (SD = 0.393).

To verify how reliable these perceptual judgments are, we calculated Cronbach's alpha (α) and intraclass correlation coefficient (ICC[1,k]). The scores on vocal attractiveness showed a high degree of internal consistency (α = 0.862) and strong interrater reliability (ICC = 0.683). The scores on affective engagement showed similar consistency (α = 0.853) and acceptable interrater agreement (ICC = 0.612) results. The ICC values here represent the variability to which one can expect these ratings to vary when affective constructs are assessed via subjective perception. The model's suitability to make precise predictions on these ratings feeds into our argument about its strength to perceive affective lessons during learning processes via these signals.

*B. Preprocessing and Feature Extraction*

This paper extracted four categories of speaker-side features, i.e., facial dynamics, oculomotor signals, acoustic prosody, and semantic content, from asynchronous instructional videos. Each modality was independently processed and aligned per video segment to construct a multimodal representation for predicting affective engagement and vocal attractiveness.

1) **Facial features**

Facial features were extracted using MediaPipe FaceMesh, which detects 478 three-dimensional facial landmarks per frame. These landmarks were grouped into seven anatomically defined facial regions (e.g., forehead, eyes, nose, lips, cheeks, jaw, and eyebrows). For each region, we computed geometric and dynamic descriptors, including the mean and standard deviation of landmark positions, inter-point distances, angular relations derived from landmark triplets, and inter-frame displacement velocity.

A fixed 2-second time window (corresponding to 60 frames at 30 fps) was used to extract temporal sequences of facial features. For each frame, a 63-dimensional region-level feature vector was generated, and the vectors were subsequently flattened and concatenated across the 60-frame window, resulting in a 3,780-dimensional facial feature vector (60 × 63) per video segment. This approach preserves short-term dynamic patterns in facial expressiveness while maintaining a consistent input shape for regression. All features were z-normalized and stored in .npy format for reproducibility [42].

2) **Oculomotor features**

Oculomotor features capture the speaker's visual attention and cognitive intent, offering complementary signals to facial dynamics. In this study, eye tracking and gaze behavior were derived from the same geometric landmark set used for facial analysis, specifically targeting eye-region landmarks (e.g., points 33, 133, 362, and 263) as defined by the MediaPipe FaceMesh model. From frame-by-frame gaze vector estimation, we computed a set of oculomotor indicators, including average fixation duration, total saccadic transition count, and statistical dispersion (mean and standard deviation) of these movements. These metrics were aggregated into a seven-dimensional feature vector per video segment and z-normalized across the dataset. Integrated into the affective engagement model, these features enabled the inference of attentional dynamics and speaker focus, enhancing model interpretability under varying expressive conditions [11].

3) **Acoustic features**

The audio signals were pre-processed following a standard operating procedure consisting of VAD, band pass filtering, noise suppression, loudness normalization, and silence trimming. To standardize vocal conditions, features were derived within a 2-second window centered within each video file. The window positioning is done utilizing a sliding window strategy with a window step size of 10 seconds across the whole audio file. The window corresponding to low prosodic variability is identified within each window by computing the combined standard deviation values within F0 and root mean square (RMS) energy.

The theoretical guidance apparent within our choice to select a 2-second segment, wherein there is a certain level of prosody stability, is to ensure a similar acoustic environment is present for each speaker. The main benefit which can be accrued within vocal periods that contain a certain amount of vocal stability is that there is a representative insight into a speaker's true voice quality, which discounts variability across speakers rather than variability within, as there could be a level of situational noise present within aperiodic vocal patterns which could compromise model generalization by introducing noise into the acoustical input.

The extraction of features is implemented by utilizing a popular Python library called Librosa 0.10.2 [43]. The list of features that this script extracts is comprised of MFCC values (40 features), Pitch (F0), formants (F1 to F5 values), chroma features (12 values), mel spectrogram (128 values), spectral contrast (7 values), Tonnetz features (6 values), HNR, jitter, shimmer, and prosody features such as RMS, CQ, DTQ, and CI. These features are generally preferred to acknowledge clarity, spectral, and prosodic qualities.

To model vocal attractiveness, a 203-D vector was created by combining all-important features such as MFCC, F0, F1 to F5, HNR, jitter, shimmer, CQ, DTQ, and CI, which were identified based on their known link to speaker attractiveness and affective expression [44].

To build the affective engagement model, a selective set that contained vital prosodic and spectral information was added to the multimodal input. The vectors were z-normalized and saved with the .npy extension.

4) **Textual features**

Audio segments were transcribed using Whisper-large-v2, a state-of-the-art automatic speech recognition (ASR) model optimized for Mandarin. The resulting transcripts were transformed into sentence-level embeddings using the multilingual paraphrase-MiniLM model from SentenceTransformer, which preserves semantic and affective context in short instructional utterances [45].

For each video segment, a 2-second alignment window was used to synchronize multimodal inputs. For textual features, the transcript corresponding to each 2-second window was extracted, and a sentence embedding was generated using the multilingual paraphrase-MiniLM model. This produced a 384-dimensional semantic vector per segment, which was then z-normalized across the dataset. These textual features were exclusively included in the affective engagement model to capture emotionally expressive language and speaker intent. They were excluded from the vocal attractiveness model, which relied solely on acoustic input [46], [47].

*C. Affective Engagement Regression Modeling*

To infer affective engagement solely from speaker-side cues, we developed a supervised regression model integrating facial, acoustic, and textual modalities. The output was a continuous score aligned with the 1–5 Likert scale used in post-video user evaluations [40].

Facial features capture temporal sequences of region-level motion dynamics, which were flattened across the 60-frame window to form a fixed-length 3,780-dimensional representation [48]. The oculomotor signals, derived from the same eye-region landmarks used in facial analysis, were treated as an independent feature stream to quantify gaze-specific attentional behavior while maintaining interpretability in the multimodal framework.

Acoustic features were extracted from a 2-second segment characterized by prosodic stability, identified as the period of minimal pitch ($F_0$) and energy fluctuation, to represent each speaker's baseline vocal profile. Semantic embeddings, averaged per video segment, served as the textual input.

Each modality was processed independently. The resulting vectors were concatenated to form a unified multimodal feature vector:

$$x_i = [f_i^{(facial)}, f_i^{(Oculomotor)}, f_i^{(acoustic)}, f_i^{(textual)}] \quad (1)$$

where $x_i$ denotes the multimodal features for the $i$-th video segment, and $f_i \in [1, 5]$ is the corresponding ground-truth engagement score. The concatenated input was fed into an XGBoost regressor trained to minimize mean squared error (MSE):

$$\mathcal{L}_{affective\_engagement} = \frac{1}{N}\sum_{i=1}^{n}(y_i - F_\theta(x_i))^2 \quad (2)$$

where $N$ denotes the number of training samples, and $F_\theta$ denotes the regression function parameterized by $\theta$.

The approach to feature-level fusion makes traceability possible between the modalities, which helps model the interaction between the two modalities. The choice of model, XGBoost, is suitable for handling high-dimensional and divergent feature spaces [51], as seen in this study, which deals with 3,780-dimensional facial representations and 203-dimensional acoustic representations. The framework has an ensemble-learning characteristic that aids in accumulating decision trees to model interactions among various features, preventing overfitting by shrinkage and subsampling, making this model particularly robust and suitable for multimodal learning applications, including multimodal affective learning, due to its strong regulation capabilities against sparse and imbalanced representations. The model can interpret features at the feature level by analyzing Shapley values, which makes it suitable for multimodal applications involving consistent interpretation, including multimodal affective learning applications.

Hyperparameters were optimized using Bayesian search, and five-fold cross-validation was employed for internal validation. To prevent information leakage, all processing was performed strictly within the train/validation/test split, including feature extraction, model training, and evaluation. The modeling pipeline was implemented using Librosa 0.10.2 for audio processing, NumPy for concatenating all extracted features, and XGBoost 1.7 in Python 3.10.

*D. Vocal Attractiveness Regression Modeling*

To predict speaker vocal attractiveness, we trained an independent XGBoost regressor using only acoustic features extracted from the prosodically stable audio segments, as described in Section III B.3. Each segment was encoded into a z-normalized 203-dimensional acoustic vector comprising F0, HNR, jitter, shimmer, MFCCs, F1–F5, and prosodic rhythm indicators such as CQ, DTQ, and CI. These features were selected based on prior work linking them to vocal clarity, emotional expressiveness, and perceived speaker likability [44].

Variations in F0 and formant dispersion have been shown to influence the perception of vocal femininity and masculinity. At the same time, measures such as jitter, shimmer, and HNR are associated with vocal appeal and affective salience [50].

Let $a_j$ denote the acoustic feature vector for the $j$-th video, and $r_j \in [1,5]$ he corresponding vocal attractiveness rating. The regression function $g_\phi$ parameterized by $\phi$ was trained to minimize MSE:

$$\mathcal{L}_{vocal\_attractiveness} = \frac{1}{M} \sum_{j=1}^{M} \left( r_j - g_\phi(a_j) \right)^2 \quad (3)$$

where $M$ is the number of training instances.

The same modeling protocol as the engagement model was applied, including Bayesian hyperparameter optimization, five-fold cross-validation, and regularization to mitigate overfitting. All preprocessing and modeling procedures were implemented using the same toolchain as detailed in Section C, ensuring methodological consistency and reproducibility.

## IV. RESULTS AND EVALUATION

Two supervised regression models were independently trained and evaluated, as described in Section III: a multimodal model for predicting affective engagement, and an acoustic-based model for estimating vocal attractiveness. Both employed XGBoost with hyperparameters optimized via Bayesian search.

For each task, the dataset was partitioned into training (70%), validation (15%), and test (15%) sets using stratified sampling to maintain the target score distribution across splits. Five-fold cross-validation was performed on the training set to guide model selection and reduce the risk of overfitting.

### A. Affective Engagement Model

XGBoost was selected for its effectiveness in high-dimensional regression tasks involving heterogeneous input sources. The affective engagement model was trained using 4,374-dimensional input vectors (D) formed by concatenating four modality-specific feature sets: acoustic (203D), oculomotor (7D), semantic (384D), and facial (3,780D). All features were z-score normalized before concatenation.

A speaker-independent partitioning strategy was employed to divide the dataset (n = 10,360) into training (n = 7,252), validation (n = 1,554), and test sets (n = 1,554). Hyperparameter tuning was performed using Bayesian optimization on the validation set, minimizing RMSE as the objective function. The optimal parameter configuration was identified as: α = 3.3977, λ = 8.8721, η = 0.1457, max_depth = 13, colsample_bytree = 0.6635, subsample = 0.6789, and γ = 0.0069.

The model converged after 164 boosting rounds. The validation RMSE stabilized at 0.129, and the test RMSE reached 0.124, indicating consistent generalization performance without signs of overfitting.

To assess the contribution of each modality, an ablation analysis was conducted by training four unimodal models using the same architecture, hyperparameter setup, and evaluation metrics. As summarized in Table I, the multimodal model outperformed all unimodal baselines across MSE, RMSE, mean absolute error (MAE), and R², highlighting the additive value of cross-modal integration.

**Table I**. Affective Engagement Model Results

| Modality | MSE | RMSE | MAE | R² |
|---|---|---|---|---|
| Facial | 0.048 | 0.220 | 0.130 | 0.681 |
| Oculomotor | 0.081 | 0.285 | 0.171 | 0.483 |
| Acoustic | 0.043 | 0.208 | 0.121 | 0.722 |
| Textual | 0.076 | 0.275 | 0.154 | 0.520 |
| Multimodal | 0.015 | 0.124 | 0.086 | 0.850 |

**Fig. 2** demonstrates a strong linear relationship between the predicted and actual engagement scores, indicating that the model reliably captures core affective patterns across diverse speaker profiles.

The attained R² of 0.85 reinforces the predictive power of speaker-side multimodal features. The superior performance of the fused model, compared to its unimodal counterparts, demonstrates that each modality contributes distinct and complementary information for estimating affective engagement.

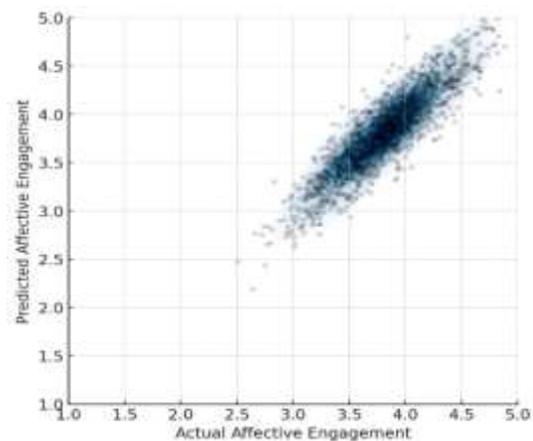

**Fig. 2**. Correlation between actual and predicted affective engagement.

### B. Vocal Attractiveness Model

The vocal attractiveness model was implemented as an independent XGBoost regression pipeline using only acoustic inputs. The dataset comprised 9,960 video samples annotated with audience-rated vocal attractiveness scores on a 1–5 scale (see Section III.A). The samples were split into training (6,972), validation (1,494), and test (1,494) sets using a speaker-independent partitioning strategy.

Each input was a 203-dimensional feature vector derived from prosodic and spectral attributes extracted from the most vocally stable 2-second segment (see Section III.B), followed by a z-score normalization. Bayesian optimization was performed on the validation set to minimize RMSE, resulting in the following optimal hyperparameters: λ = 7.9300, η = 0.0630, max_depth = 9, γ = 0.0500, colsample_bytree = 0.9800, and subsample = 0.7400.

The model converged after 372 boosting rounds. The training RMSE reached 0.0479, and the validation RMSE stabilized at 0.158, indicating no evidence of overfitting based on validation performance. This acoustic-only model provides a scalable affective prediction approach suitable for privacy-sensitive or audio-restricted scenarios. Table II summarizes model performance on the held-out test set.

**Table II**. Vocal Attractiveness Model Results

| Modality | MSE | RMSE | MAE | $R^2$ |
|---|---|---|---|---|
| **Acoustic** | 0.021 | 0.144 | 0.098 | 0.881 |

**Fig. 3** illustrates the close correspondence between predicted and actual vocal attractiveness ratings, with the diagonal pattern indicating reliable capture of perceived vocal appeal. Accounting for 88% of the variance in listener ratings, the results highlight the potential of acoustic features as reliable indicators of perceived vocal attractiveness in asynchronous learning scenarios.

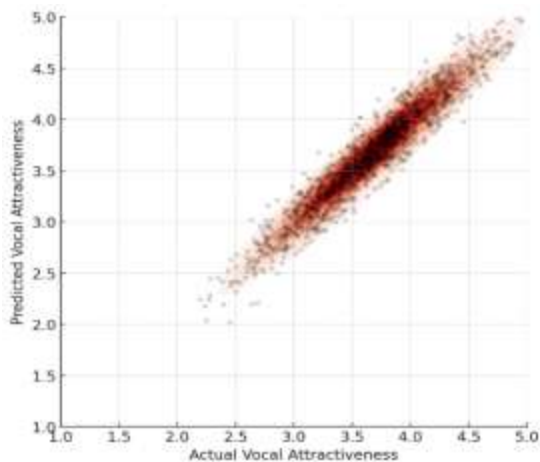

**Fig. 3**. Correlation between actual and predicted vocal attractiveness.

### C. Role of Vocal Attractiveness in Engagement Prediction

To investigate the link between vocal attractiveness and affective engagement, we first computed the Pearson correlation between human-rated scores in the test set (n = 1,494). A strong positive correlation was observed (r = 0.732, p < .001), suggesting that enhanced perceived vocal attractiveness positively correlates with affective engagement. Similarly, the predicted scores from the two models exhibited a comparably strong correlation (r = 0.743, p < .001), suggesting consistent alignment between acoustic salience and inferred affective outcomes.

This finding reinforces previous research on the role of vocal prosody, such as pitch modulation, clarity, and expressive intonation. The finding shapes perceived speaker likability and emotional resonance [5], [40]. Although the affective engagement model integrates multiple modalities, its close alignment with the acoustic-only vocal attractiveness model highlights the standalone predictive strength of prosodic features.

Significantly, as shown in Table I, the acoustic-only model outperformed all other unimodal variants, achieving an $R^2$ of 0.722, second only to the multimodal model ($R^2$ = 0.850). This emphasizes the centrality of the acoustic stream in engagement inference and supports its utility in scenarios where visual or textual data may be unavailable due to privacy, bandwidth, or quality constraints.

**Fig. 4** presents this relationship via a scatter plot with a fitted regression line. The apparent linear trend across the 1–5 scale illustrates that vocal attractiveness can function as a practical proxy at the predictive level in engagement estimation pipelines. From a system design perspective, the dual-model architecture enhances robustness and applicability. When visual or textual data are unavailable, due to privacy constraints or data sparsity, the acoustic-based model alone can still yield reliable engagement estimates. This modular design supports flexible deployment across varied digital learning environments, aligning with real-world constraints.

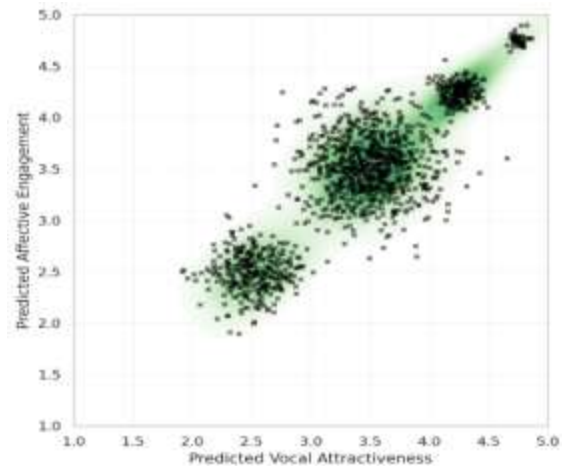

**Fig. 4**. Correlation between predicted vocal attractiveness and affective engagement.

Using SHAP (Shapley Additive Explanation) values, we can analyze and explain both models on how much each cue in expressive speech is contributing to each model's predictions. The vocal attractiveness model (with an $R^2$ value = 0.88) shows that audience perception concerning vocal clarity and control is governed by such acoustical factors as $F_0$, HNR, excessive jitter, and shimmer, which negatively contribute to how attractive one's voice is. The affective engagement model (with an $R^2$ value = 0.85) shows that prosody variability, facial dynamics around regions of eyes and mouth, and oculomotor control positively contribute to audience engagement. Shared prosody features can serve as a link between both models because communicating vocal attractiveness is a summary acoustic property that is related to audience engagement, and this property is enhanced by visual features added by multimodal models.

Despite the relatively low interrater agreement on affective engagement (ICC = 0.612) and vocal attractiveness (ICC = 0.683) being merely moderate, this degree of variability is to be expected because affective and aesthetic perception necessarily involve subjective experiences. The high degree of prediction precision is thus not inconsistent with these figures but rather suggests that these models tap into this collective perceptual consensus (that is, human judgment's center of agreement) while suppressing variability across raters specific to each. This finding is consistent with these models being sensitive to those aspects most generally similar across human subjects, those aspects involved in affective engagement, which correspond to building scalable predictors of collective perception.

## V. CONCLUSION

This paper presents a privacy-conscious Emotion AI framework that infers audience-perceived affective engagement and vocal attractiveness from asynchronous instructional videos using only speaker-side multimodal cues. Leveraging a large-scale MOOC dataset and XGBoost regression, the proposed dual-model system achieved strong predictive performance ($R^2 = 0.85$ and $0.88$) without relying on audience-side data. Notably, acoustic features alone explained 72% of the variance in engagement, supporting their standalone predictive value. The modular, feature-level design enhances interpretability, scalability, and alignment with AI governance standards such as the EU AI Act. Besides being useful as a professional solution in MOOC environments, this study offers the first possible privacy-friendly benchmark applicable to multimodal engagement model prediction relying solely on speaker-side patterns, constituting a foundation stone toward future research developments related to building scalable emotion-based learning analytic tools as well as virtual coaching systems described as privacy-friendly computational social systems.

## ACKNOWLEDGMENT

This work was supported by Taiwan's National Science and Technology Council (NSTC) under grants 113-2622-H-003-003 and 112-2410-H-003-102-MY2. The authors thank the partnering MOOC platforms and participants for their contributions.

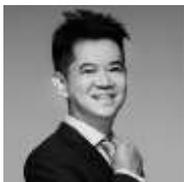
**Hung-Yue Suen** received the Ph.D. degree in Management Information Systems from National Chengchi University, Taipei, Taiwan. He is currently a Full Professor in the Department of Technology Application and Human Resource Development at National Taiwan Normal University, Taipei, Taiwan. His research interests include computational social systems, human–AI interaction, and AI applications in talent assessment.

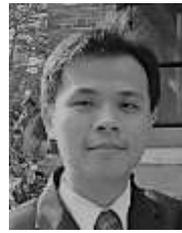
**Kuo-En Hung** received the Ph.D. degree in Human Resource Technology from National Taiwan Normal University, Taipei, Taiwan, and the M.S. degree in Computer Science from National Tsing Hua University, Hsinchu, Taiwan. He is currently an industrial researcher at National Taiwan Normal University. His research interests include computer vision, acoustic modeling, neural networks, and social computing.

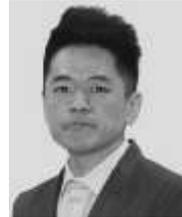
**Fan-Hsun Tseng** (Senior Member, IEEE) received the Ph.D. degree in Computer Science and Information Engineering from the National Central University, Taoyuan, Taiwan, in 2016. In 2021, he joined the faculty of the Department of Computer Science and Information Engineering, National Cheng Kung University, Tainan, Taiwan, where he is currently an Associate Professor since 2024. His research interests include federated learning, reconfigurable intelligent surfaces, and intelligent reflecting surfaces. He has served as the Director and the Chair of the Computer Society Chapter of the IEEE Tainan Section. He is a senior member of the IEEE and of ACM.